\documentclass[epj,final]{svjour}

\usepackage{latexsym}
\usepackage{url}
\usepackage{amsfonts}
\usepackage{amssymb}

\RequirePackage{graphicx}

\begin{document}
\title{Two fundamental constants of gravity unifying the dark matter and the dark energy}
\author{V.G. Gurzadyan
, A. Stepanian
}                     
%
%
\institute{Center for Cosmology and Astrophysics, Alikhanian National Laboratory and Yerevan State University, Yerevan, Armenia}
\date{Received: date / Revised version: date}
%

\abstract{
The common nature of the dark sector - dark energy and dark matter - as shown in \cite{G17} follows readily from the consideration of generalized Newtonian potential as a weak-field General Relativity. That generalized potential satisfying the Newton's theorem on the equivalence of sphere's gravity and that of a point-mass located in its center, contains an additional constant which along with the gravitational constant is able to explain quantitatively both the dark energy (cosmological constant) and dark matter. So, gravity is defined not by one but two fundamental constants. We show that, the second constant is dimensional-independent and matter-uncoupled and hence is even more universal than the gravitational constant, thus affecting the strategy of observational studies of dark energy and of the search of dark matter.    
} 

%
\maketitle
\section{Introduction}

The discovery of the dark sector as a dominant constituent of the Universe is one of outstanding recent astrophysical achievements and continues to be a key puzzle for physical theories. Various modifications of the Newtonian gravity and of General Relativity (GR) are being actively considered in that context.

Among the possible approaches to the modified gravity, including GR, is the one based on a theorem proved by Newton in "Principia" on the equivalence of the gravity of sphere and that of a point mass located in its center. The principal importance of that theorem was obvious, since the motion of the planets which were spheres and not point masses, could be considered explained by gravity law only upon the proof of that theorem. Now, it appears that this theorem provides a two-step path to modified gravity theories and directly to the dark sector problem \cite{G17,G85}:

1. The general function satisfying that theorem provides an additional term containing a constant and thus modifying the Newtonian gravity;

2. That modified Newtonian gravity leads to a modified GR with the former as its weak-field limit.  

So, the modified GR has to initially include that additional constant, along with the gravitational constant. As shown in  \cite{G17,G85}
that additional constant entering both the modified Newtonian gravity and GR enables to describe by its sign and quantitative value both the dark matter and the dark energy. That constant appears to be the renown cosmological constant which was introduced by Einstein \cite{E} in order to have static solutions to Einstein equations. 

The fact that the two constants of the gravitational interaction are able to describe self-consistently, i.e. without postulation of additional scalar or other fields, the dark matter and dark energy, reveals their unified, gravitational nature. Namely, the dark matter appears as a result of pure gravitational interaction, but with a law containing the additional constant.

We analyze this approach and reveal the role of that additional universal physical constant in the classical and relativistic gravities. The consequences of this approach can have direct impact on the strategy of the observational studies of the dark energy and the search of the dark matter.

\section{Newton's theorem and General Relativity}

The general function to satisfy the Newton's theorem that a sphere acts as a point-mass located in its center has the following form for the force \cite{G85} 
\begin{equation}\label{form}
f(r)= C_1 r^{-2} + C_2 r,
\end{equation}   
where $C_1$ and $C_2$ are constants and $f(r)$ is the solution of equation
\begin{equation}
\frac{r^2}{2}f''(r) + r f'(r)-f(r)=0.
\end{equation}

Thus, one can conclude that according to Newton's theorem the gravitational force should contain two terms i.e. an inversed square term and a linear one.

Considering the original formulation of gravity by Newton himself, it becomes clear that the constant $C_1$  is written as $C_1=G\, m^2$, i.e. with the familiar gravitational constant $G$ and mass $m$, the latter entering also the law of mechanics. In this sense, the Newtonian gravity can be regarded as a very special case i.e. $C_2=0$ of all possible forms of gravitational fields where one can consider spherical objects as points. Furthermore it should be noticed that, Newton himself did not consider the most general form of the force before formulation of his theory of gravity, although he proved that in the context of his theory it is possible to consider spheres as points. 

In this context the presence of a linear term was forgotten for hundreds of years until the formulation of GR and introduction of cosmological constant $\Lambda$ 
\begin{equation}
G_{\mu\nu} + \Lambda g_{\mu\nu}= \kappa T_{\mu\nu}.
\end{equation}

After that it became obvious that, by considering the $\Lambda$, as introduced above, the GR's weak field limit will contain an additional linear term. In this sense the metric tensor components for the sphere's gravity in the weak-field limit will be
\begin {equation} \label {GRC}
g_{00} = 1 - \frac {2 G m}{r c^2} - \frac{\Lambda r^2}{3};\,\,\, g_{rr} = 1 + \frac{2 G m}{r c^2} + \frac {\Lambda r^2}{3}.
\end {equation} 

So, one can conclude that although for the first time the existence of $\Lambda$ was proposed by Einstein in the context of GR \cite{H}, it would be possible to find the full GR equations (Eq.(3)), if Newton had considered both terms and formulated his theory based on Eq.(1). In this sense, by considering the Newton's principle and the most general form of the force, the cosmological term appears in Einstein's equations not by principles of GR, but as the second linear term of Newtonian gravity.

It should be noticed that, although Eq.(4) has been considered previously in different contexts (e.g. \cite{H} and \cite{Bah} and references therein), the approach of \cite{G17,G85} from the roots of Newtonian gravity/GR provides an insight to the unified nature of the dark matter and the dark energy. Namely, the presence of $\Lambda$ as an additional linear term in Newtonian regime i.e. Eq.(4) enables one to describe the dark matter in galaxies, as the cosmological constant in GR describes the dark energy, and as shown in \cite{G17} both values of $\Lambda$, i.e. those describing the dark matter and dark energy (cosmological constant) quantitatively agree with each other. In the case of the dark matter, it is of principal importance since Eq.(1) describes a non-force-free field inside a shell except its center, while for Newtonian law the force-free field is entirely inside the shell. This fact agrees with the observational evidence that the galactic halos determine the properties of galactic disks \cite{Kr}.

\section{Group-theoretical analysis of Newton's theorem}

In previous section we have shown that, it is possible to justify the existence of second term in Eq.(1) as the weak-field limit of GR equations written with $\Lambda$. However, as mentioned above $\Lambda$ was introduced not by Newton's theorem but according to conservation of Energy-Momentum tensor and the fact that $\partial^{\mu}g_{\mu\nu}=0$. So it seems quite reasonable that, to make a more powerful justification, we try to infer the Newton's theorem based on above relativistic considerations. Thus we turn to the isometry groups.    

In Eq.(3), depending on $\Lambda$'s sign  - positive, negative or zero -  one has three different vacuum solutions (three different asymptotic limits) for the field equations as shown in Table \ref{tab1}.

\begin{table}
\caption{}\label{tab1}
\begin{tabular}{ |p{2cm}||p{4cm}|p{4cm}|p{2cm}| }
\hline
\multicolumn{4}{|c|}{Background geometries for vacuum solutions} \\
\hline
Sign& Spacetime&Isometry Group&Curvature\\
\hline
$\Lambda > 0$ &de Sitter (dS) &O(1,4)& +\\
$ \Lambda = 0$ & Minkowski (M) & IO(1,3) &0\\
$\Lambda <0 $ &Anti de Sitter (AdS) &O(2,3)& -\\
\hline
\end{tabular}
\end{table}
The interesting feature of all these 4-dim maximally symmetric Lorentzian geometries is that, for all of them the stabilizer subgroup of isometry group is the Lorentz group O(1,3). This means that at each point of all these spacetimes, one has an exact Lorentz symmetry. Since O(1,3) is the group of orthogonal transformations, one can conclude that all above spacetimes possess spherical symmetry (in Lorentzian sense) at each point. Speaking in terms of geometry, for above three spacetimes we have
\begin{equation}\label{qspace}
dS= \frac{O(1,4)}{O(1,3)}, \quad M=\frac{IO(1,3)}{O(1,3)}, \quad AdS=\frac{O(2,3)}{O(1,3)}.
\end{equation}
It is clear that in non-relativistic limit the full Poincare group IO(1,3) is reduced to Galilei group Gal(4)=(O(3)$\times$R)$\ltimes$R$^6$, which is the action of O(3)$\times$R (as the direct product of spatial orthogonal transformations and of time translation) on group of boosts and spatial translations R$^6$. In the same way one can find the non-relativistic limit of O(1,4) and O(2,3) groups
\begin{equation}\label{nrel}
O(1,4) \to (O(3) \times O(1,1)) \ltimes R^6,\quad O(2,3) \to (O(3) \times O(2)) \ltimes R^6.
\end{equation}
Furthermore, considering the fact that the Galilei spacetime is achieved via quotienting Gal(4) by O(3)$\ltimes$R${}^{3}$ (the group generated by orthogonal transformations and boosts), one can continue the analogy and find the so-called Newton-Hooke NH(4)${}^{\pm}$ spacetimes by same quotient group but now for groups of Eq.(\ref{nrel}) (see \cite{NH1}\cite{NH2}\cite{NH3}). In this sense, depending on the sign of $\Lambda$, we can not only find the general form of the Newtonian modified gravity (according to section 2), but also the non-relativistic background geometries of the Lorentzian spacetimes in Table \ref{tab1} and their symmetries.

To complete the proof, one has to check whether it is possible to apply the Newton's theorem to these spacetimes or not. As stated above, to apply the gravity law to planets (spheres) Newton considered them as points. Speaking in terms of mathematics it means that at each point one should have O(3) symmetry. This statement is similar to what we showed for 4-dim geometries of Table \ref{tab1} and the Lorentz group O(1,3). The possible 3-geometries with such property are listed in Table \ref{tab2}.
\begin{table}
\caption{}\label{tab2}
\begin{tabular}{ |p{4cm}||p{4cm}|p{4cm}|p{2cm}| }
\hline
\multicolumn{3}{|c|}{3D Background Geometries with O(3) as the stabilizer} \\
\hline
Spacetime&Isometry Group&Curvature\\
\hline
Spherical &O(4)& +\\
Euclidean & E(3) &0\\
Hyperbolic &O${}^{+}$(1,3)& -\\
\hline
\end{tabular}
\end{table}

Recalling that for non-relativistic theories we have two absolute notions of space and time geometry (in contrast to relativistic theories where space and time are unified in spacetime geometry), the last step is to check whether the spatial geometry of two NH(4)${}^{\pm}$ spacetimes, as well as the Galilei spacetime are equal to one of the geometries mentioned in Table \ref{tab2} or not. There are several ways to check this statement, however the most straightforward one is to check the algebraic structure of spatial geometry. Recalling the fact that for both NH(4)${}^{\pm}$ spacetimes and Galilei spacetime the spatial algebra is identical and equal to Euclidean algebra E(3)$=$R${}^3$$\rtimes$ O(3), we can conclude that for all above spacetimes we have an exact O(3) symmetry at each point of spatial geometry. In this sense we will arrive at Newton's theorem based on group theoretical analysis of GR equations.

\section{Newton's theorem in d-dimensions}

To throw more light on the constant $\Lambda$ we consider the higher dimensional cases which simply means that the gravitational field defined on $S^{d-1}$ should be equal to that defined for a single point at d-dimensional space. For the potential one has 
\begin{equation}\label{GenPrin}
\Delta_{S^{d-1}} \Phi=C_1,
\end{equation}
where $\Delta_{S^{d-1}}$ denotes the Laplace operator defined on $S^{d-1}$ and the constant $C_1$ defines the mathematical feature of geometrical point.
Now due to spherical symmetry we can write 
\begin{equation}\label{GenPrinr}
\frac{1}{r^{d-1}}(\frac{d}{dr}r^{d-1}\frac{d}{dr}\Phi)=C_1.
\end{equation}
So the most general form of the gravitational potential $\Phi$ of sphere in d-dimensional case according to Newton's theorem is 
\begin{equation}\label{GenPot}
\Phi(r)=C_1 \frac{r^2}{2d}+\frac{C_2}{(d-2)r^{d-2}}.
\end{equation}
In this equation $C_1$ is the constant of Eq.(\ref{GenPrin}) and the constant $C_2$ arises during solving the equation. Note, that for $d=2$ the second term becomes logarithmic but the first one remains unchanged.

The potential $\Phi$ in Eq.(\ref{GenPot}) at $d=3$ is not only in full agreement with Eq.(1), but also leads to further insights.  One can identify the $C_2$ in Eq.(\ref{GenPot}) with the $\Lambda$-constant at the d-dim generalization of ordinary Newtonian gravity, 
\begin{equation}\label{PotN}
\Phi(r)=-\frac{G_{d} M}{r^{d-2}}-\frac{\Lambda c^2 r^2}{2d},
\end{equation}
where $G_{d}$ indicates the d-dimensional gravitational constant. 

Note a remarkable fact: comparing the two constants - the gravitational constant $G$ and the $\Lambda$ - one can see their essential difference. Namely, the gravitational constant $G$ is dimensional-dependent and couples to matter, while $\Lambda$ is neither dimensional-dependent nor matter-coupled. Such universality of $\Lambda$ can be considered as fitting its vacuum content noticed by Zeldovich from completely different principles \cite{Z}. 
  
So, the gravity has not one, but two fundamental constants - $G$ and $\Lambda$ - and the second one (cosmological constant) is more universal (dimensional-independent) than the gravitational constant! The two constants together are able to explain quantitatively the dark energy and the dark matter \cite{G17}.   
  
Then, the metric component of $d+1$ dimensional spacetime is
\begin{equation}\label{GRCd}
g_{00}=1-\frac{2 \Phi}{c^2}.
\end{equation}
From the d-dimensional Gauss's law 
\begin{equation}\label{Gauss lawd}
\Delta \Phi= \frac{2 \pi^ {\frac{d}{2}}}{\Gamma (\frac{d}{2})} G_d \rho - \Lambda c^2,
\end{equation}
where $\rho$ is the d-dimensional density of matter. Consequently one gets the Einstein constant
\begin{equation}
\kappa_d= \frac{ 4\pi^ {\frac{d}{2}}}{\Gamma (\frac{d}{2})} \frac{G_d}{c^4}.
\end{equation}
This completes the generalization of Newton's theorem to arbitrary dimension and its correspondence to classical and relativistic theories of gravity. 

Then, for the 3 possible maximally symmetric $(d+1)$-dimensional spacetimes defined by the value of $\Lambda$ one has the following geometries
\begin{equation}\label{qspaced}
dS_{d+1}= \frac{O(1,d+1)}{O(1,d)}, \quad M_{d+1}=\frac{IO(1,d)}{O(1,d)}, \quad AdS_{d+1}=\frac{O(2,d)}{O(1,d)},
\end{equation}
as the generalizations of Eq.(\ref{qspace}); for $d=3$ one easily recovers the 4-dimensional results. It is clear that in such case, irrespective which geometrical spacetime is considered, one has exact O(1,d) symmetry at each point, which in it's turn indicates the existence of spherical symmetry of Lorentzian geometry for all points. Fixing the relativistic geometries and symmetries one easily finds their non-relativistic limits
\begin{equation}\label{nreld}
O(1,d) \to (O(d) \times O(1,1)) \ltimes R^{2d},\quad O(2,d) \to (O(d) \times O(2) ) \ltimes R^{2d}, \quad IO(1,d) \to (O(d) \times R) \ltimes R^{2d}.
\end{equation}
As in section 3, one can find the non-relativistic background geometries for each case by quotienting O(d)$\ltimes$ R${}^d$ for all three symmetric groups. The resulting spacetimes are Gal(d+1), NH${}^{+}$(d+1), NH${}^{-}$(d+1), and clearly at $d=3$ one obtains the classical spacetimes. As we have mentioned earlier, the interesting feature of these non-relativistic geometries is the fact that, in contrast to relativistic case they are not metric geometries because they do not admit single metric structure and their properties can be studied via corresponding affine connection. Furthermore from geometrical point of view, for all these three cases the spatial geometry seems to be Euclidean and the pure spatial algebra is equal to Euclidean algebra E(d). Then, since E(d)$=$R${}^d$$\rtimes$ O(d), one easily concludes that in the spatial geometry the O(d) is the stabilizer group, which in it's turn means that all points can be considered as d-dimensional spheres S${}^{d-1}$. This proves that for all these three geometries the Newton's theorem is hold. However, as mentioned above, the spatial part of all three geometries is equal to each other and the question is, how $\Lambda$ affects these geometries. The answer becomes clear if one considers the temporal parts of Eq.(\ref{nreld}). Indeed, the sign of $\Lambda$ indicates that we are living either in oscillating NH(d+1)${}^{-}$, flat Gal(d+1) or expanding NH(d+1)${}^{+}$ universe. One can also check that for all these cases, depending on the sign and the value of $\Lambda$, the affine connection can be flat, for Gal(d+1) case, and either positive or negative for NH(d+1)${}^{+}$ and NH(d+1)${}^{-}$, respectively.

To conclude this brief but principal discussion, we write down the d-dimensional ($d\neq2$) Schwarzschild metric for non-zero $\Lambda$
\begin{equation}\label{SLd}
ds^2 = (1-\frac{2 G_d M}{r^{d-2} c^2} - \frac {\Lambda r^2}{3})c^2 dt^2 - (1-\frac{2 G_{d} M}{r^{d-2} c^2} - \frac {\Lambda r^2}{3})^{-1} dr^2 -r^{2}d\Omega^2_{d-1}.
\end{equation}
Although d-dimensional cases have been considered before, our approach to GR and its weak-field limit justifies the consideration of point-like dynamics for higher dimensional spheres based on Newton's original theorem.

\section{Conclusions}

Thus, according to our analysis: 

1. Gravity has not one but two fundamental constants, the gravitational constant $G$ and an additional one, $\Lambda$, which appears readily in General Relativity with weak-field limit as modified Newtonian gravity. Moreover, the $\Lambda$-constant (the cosmological constant) is dimensional-independent and matter-uncoupled and hence can be considered as even more universal than the gravitational constant $G$;

2. The $\Lambda$-constant of gravity emerges from Newton's theorem on the identity of the sphere's gravity and that of the point-mass located in its center; 

3. Both constants, $G$ and $\Lambda$, jointly are able to explain quantitatively the dark energy and the dark matter \cite{G17}, which hence appear as gravity effects.

Also, the AdS spacetime of AdS/QFT emerges here readily from the genuine structure of classical and relativistic gravities.
Positive $\Lambda$-constant is an essential condition in Conformal Cyclic Cosmology \cite{P,GP}. 
     
The accuracy of the current tests of GR (e.g. \cite{Lares}) are still far to probe the modified gravity discussed above, however, for example, the astronomical observations of galactic halos \cite{G2} can be efficient in testing the predictions regarding the dark matter nature.

AS acknowledges the ICTP Affiliated Center program AF-04 for financial support.

\end{document}